# Direct Observation of Gate-Tunable Dark Trions in Monolayer WSe$_2$

Zhipeng Li[1#], Tianmeng Wang[1#], Zhengguang Lu[2,3], Mandeep Khatoniar[4,5], Zhen Lian[1], Yuze Meng[1], Mark Blei[6], Takashi Taniguchi[7], Kenji Watanabe[7], Stephen A. McGill[2], Sefaattin Tongay[6], Vinod M. Menon[4,5], Dmitry Smirnov[2], Su-Fei Shi[1,8]*

1. Department of Chemical and Biological Engineering, Rensselaer Polytechnic Institute, Troy, NY 12180.
2. National High Magnetic Field Lab, Tallahassee, FL, 32310.
3. Department of Physics, Florida State University, Tallahassee, Florida 32306.
4. Department of Physics, City College of New York, City University of New York,160 Convent Ave, New York, NY 10031, USA.
5. Department of Physics, The Graduate Center, City University of New York,365 5th Ave, New York, NY 10016, USA.
6. School for Engineering of Matter, Transport and Energy, Arizona State University, Tempe, AZ 85287.
7. National Institute for Materials Science, 1-1 Namiki, Tsukuba 305-0044, Japan.
8. Department of Electrical, Computer & Systems Engineering, Rensselaer Polytechnic Institute, Troy, NY 12180.
[#] These authors contributed equally to this work
[*] Corresponding author: shis2@rpi.edu

**Abstract**

Spin-forbidden intravalley dark exciton in tungsten-based transition metal dichalcogenides (TMDCs), owing to its unique spin texture and long lifetime, has attracted intense research interest. Here, we show that we can control the dark exciton electrostatically by dressing it with one free electron or free hole, forming the dark trions. The existence of the dark trions is suggested by the unique magneto-photoluminescence spectroscopy pattern of the boron nitride (BN) encapsulated monolayer WSe$_2$ device at low temperature. The unambiguous evidence of the dark trions is further obtained by directly resolving the radiation pattern of the dark trions through back focal plane imaging. The dark trions possess binding energy of ~ 15 meV, and they inherit the long lifetime and large *g*-factor from the dark exciton. Interestingly, under the out-of-plane magnetic field, dressing the dark exciton with one free electron or hole results in distinctively different valley polarization of the emitted photon, a result of the different intervalley scattering mechanism for the electron and hole. Finally, the lifetime of the positive dark trion can be further tuned from ~ 50 ps to ~ 215 ps by controlling the gate voltage. The gate tunable dark trions ushers in new opportunities for excitonic optoelectronics and valleytronics.

**Main Text**

Due to the reduced screening and enhanced Coulomb interactions, monolayer TMDCs host exciton with large binding energy on the order of hundreds of meV[1–4]. Owing to the lack of inversion symmetry and the three-fold rotation symmetry, the excitons in TMDCs possess the valley degree of freedom which can be accessed through left or right circularly polarized light, which selectively excites the excitons at the K or K' valley (corners of the Brillouin zone)[5–12]. The large spin-orbit coupling induced valence band splitting ensures the spin-valley locking[2,13–15], which results in the robust valley polarized photoluminescence (PL) even at room temperature[16–19]. However, the excitons are short-lived, with a typical lifetime in the range of a few to tens of picoseconds[20–22], which significantly limits potential applications.

Recently, it has been discovered that tungsten-based TMDCs such as $WSe_2$ and $WS_2$ have a unique bandstructure, in which the spin-orbit coupling also induces the splitting of the conduction band[23–25]. The resulted ground state of the exciton, however, is a spin-triplet state which prevents direct recombination, hence referred to as the spin-forbidden dark exciton[22,25–28]. The existence of such a long-lived dark exciton has been revealed through the application of an in-plane magnetic field[25], coupling to a plasmonic substrate[28], directly observing PL from the side[27], or collecting PL signals with an objective of large numerical aperture (N.A.)[29–32].

The spin-forbidden dark exciton, nonetheless, can also radiate through a finite out-of-plane dipole moment, which does not obey the same valley physics as the bright exciton, which arises from the combination of the inversion symmetry breaking and three-fold rotation symmetry that restricts the in-plane dipole radiation.[28,33]. As a result, we previously found that the valley-resolved PL spectra under an out-of-plane magnetic field exhibit a unique "cross" pattern in the intrinsic regime, distinctively different from other excitonic complexes[29–31]. Here we apply magneto-PL spectroscopy to investigate a top-gated monolayer $WSe_2$ device and found a similar pattern in the n- or p-doped $WSe_2$. We attribute this pattern to the dark trions, a dark exciton bound to a free electron or hole. And the existence of the dark trions is unambiguously demonstrated by directly resolving the radiation angle of the dark trions through back focal plane imaging. The binding energy of the dark trions, determined by the PL peak position difference between the dark trions and dark exciton, is found to be ~ 15 meV. Interestingly, we found the asymmetric valley polarization behavior of the positive and negative trions in the presence of the out-of-plane magnetic field, originating from the different intervalley scattering mechanism of the electron and hole. The dark trions also inherit the large *g*-factor of the dark exciton, ~ -9, leading to a large Zeeman splitting of ~ 9.0 meV under the magnetic field of 17 T. The direct observation and improved understanding of the dark trions pave the way of exploiting the charged dark exciton for optoelectronic and valleytronic applications.

The boron nitride (BN) encapsulated monolayer WSe₂ device is fabricated through a pickup method as described in previous works[29,34–36], and it is shown schematically in Fig. 1a. The monolayer WSe₂ is contacted by a few-layer graphene electrode, and is also gated through a few-layer graphene as the top gate electrode, with the top BN flake working as the gate dielectric. The high quality of monolayer WSe₂ is demonstrated through the well-resolved excitonic fine structures in the gate-dependent PL spectra at 4.2 K (Fig. 1b and 1c). Most of the excitonic complexes have been identified previously, including the bright exciton ($X_0$), the biexciton (XX) and charged biexciton ($XX^-$)[29–32,37–39], the positive bright trion ($X^+$), the two negative trions ($X_1^-$ and $X_2^-$)[8,23,40,41] and the dark exciton replica ($X_D^R$)[42]. In particular, the PL peak centered at 1.690 eV, with the intensity maximized in the nearly charge neutral region (gate voltage between ~ -1.0 V to ~ 0.35 V), has been identified as the spin-forbidden dark exciton ($X_D$) previously[22,25–28,33]. The spin-forbidden $X_D$, schematically shown in Fig. 1d, arises from the unique bandstructure of WSe₂, in which the spin-orbit coupling induced conduction band splitting leads to a conduction band minimum (CBM) with the spin opposite to that of the valence band maximum (VBM). As a result, the ground state of the optical excitation generated exciton is a spin-triplet state and its direct recombination is spin forbidden.

The valley-resolved magneto-PL spectra have been previously employed to identify the dark exciton in charge-neutral WSe₂. When the WSe₂ is excited with a circularly polarized light (say $\sigma^-$), the radiation from the out-of-plane dipole is p-polarized and will give rise to both left ($\sigma^-$) and right ($\sigma^+$) circularly polarized PL emission in our detection scheme, which have different energies under the out-of-plane magnetic field, a result of the opposite Zeeman splitting for the exciton in K and K' valleys[16–19,37,43–46]. Therefore, for charge-neutral WSe₂, the valley resolved magneto-PL spectra in the $\sigma^-\sigma^-$ configuration (SI), standing for the left-circular polarized excitation and left-circular polarized detection, show that the PL peak corresponding to $X_D$ splits into two branches, while the other excitonic complexes, such as $X_0$, exhibit a linear Zeeman shift. In this study, we found that as we gate the WSe₂ to the hole-doping (gate voltage -2.0 V) and electron-doping regime (gate voltage 1.0 V), the PL peaks centered at ~1.675 eV and 1.674 eV with no magnetic field applied, labeled as $X_D^+$ and $X_D^-$ in Fig. 1b, exhibit similar "cross" patterns in the valley-resolved magneto-PL spectra shown in Fig. 2a and 2b, respectively. We have also reproduced this observation in two other BN encapsulated WSe₂ devices (see the Supplementary Note 1).

The positive dark trion and negative dark trion, schematically shown in Fig. 1e and 1f, emit photons through the recombination of the electron and hole either in the K or K' valley, same as the dark exciton (Fig. 1d). As a result, in a non-interacting picture in which the Zeeman splitting is calculated through the individual shift of CBM and VBM associated with the recombination electron-hole pair[29,47], the magneto-PL spectra of $X_D^+$ and $X_D^-$ will share the same Zeeman splitting as the charge-neutral $X_D$. Therefore, we assign the emerging "cross" patterns in the hole-doped and electron-doped WSe₂ (Fig. 2a and 2b) to the positive dark trion and the negative dark trion, a neutral dark exciton bound to a free hole and an electron, respectively. We notice that a very recent report also utilizes the similar pattern in the magneto-PL spectra to identify the dark trions[48].

This determination is also supported by the quantitative analysis of the Zeeman splitting[16–19,28,35–38]. The valley-Zeeman splitting can be expressed as: $E^{K(K')} = E_0 \pm g\mu_B B$, where $E_0$ is the PL peak position at the absence of the external magnetic field, "+" or "-" corresponds to the shift in the K or K' valley, $g$ is the Landé $g$-factor, $\mu_B$ is the Bohr magneton. The Landé $g$-factor, as previously mentioned, can be calculated by counting the difference of the shift of the conducting band minimum (CBM) and valence band maximum (VBM) that are associated with the electron and hole involved in the recombination. The $g$-factor for $X_D$ in this device is ~ -9.5 (SI), consistent with the theoretical expectation of -8 and previous experimentally reported value[22,29,31]. Fig. 2c and 2d show that the $g$-factors of $X_D^+$ and $X_D^-$ are ~ -8.6 and ~ -9.8, respectively, similar to that of $X_D$, considering the fitting uncertainty, experimental uncertainty and possible change of the $g$-factors with the increased doping level. According to our discussion, these agreements support the assignments of the positive dark trion and the negative dark trion, as the recombination e-h pairs involved in the dark trion is the same as the charge-neutral dark exciton. The g-factor calculation is also confirmed from the data from other two devices (see Supplementary Note 2 and Supplementary Note 3 for the data from the second and third devices). The $g$-factors of $X_D^+$ and $X_D^-$, similar to that of $X_D$, are more than two times larger than that of $X_0$ (~ -4.3)[19,29,49,50]. As a result, the Zeeman splitting of the dark trions can be as large as ~ 9.0 meV at the presence of an out-of-plane magnetic field of 17 T. The significant Zeeman splitting due to the large g-factor can be utilized for breaking the valley degeneracy.

Despite the similar binding energy, the negative and positive dark trions exhibit asymmetric behavior. The intensity of the negative dark trion is weaker than that of the positive dark trion (Fig. 1b and 1c), consistent with previous reports[28,51]. It is likely due to the existence of the lower energy states in the n-doped WSe$_2$, such as the one suggested by the bright PL peak near energy 1.660 eV in the n-doped WSe$_2$ (Fig. 1b). It is also worth noting that due to the weak PL intensity of the negative dark trion, the "cross" pattern is more difficult to observe in the n-doped WSe$_2$ (see Supplementary Note 2 and 3). More interestingly, the splitting K and K' branches PL in the magneto-PL spectra are not of equal intensity for dark trions, in contrast to the case of the dark exciton[29,42] that has equal emission probability form K and K' branches under the external out-of-plane magnetic field. In addition, the brighter branch of the negative dark trion ($X_D^-$) (with the intensity ratio ~ 20 for the PL intensity between the two branches) is the same as the branch visible in the magneto-PL for the bright exciton ($X_0$), but the opposite to the brighter branch of the positive dark trion ($X_D^+$) (with the intensity ratio ~ 4). This means that dressing the dark exciton with one free hole or electron makes the dark trion to emit light more likely from one valley or the other, granting the valley information to the dark trion.

This information can be quantitatively shown in Fig. 3. Since only one branch of the bright exciton is visible in the helicity-resolved PL spectra under the out-of-plane magnetic field, we define the valley polarization of the bright exciton under the magnetic field as $P_B$=1. $P_B$ is defined as $P_B = \frac{I(K')-I(K)}{I(K')+I(K)}\sigma(B)$, where *I(K')* and *I(K)* are the PL intensities for K' valley and K valley. $\sigma(B)$ is 1 or-1 for positive or negative B field, respectively. It is evident from Fig. 3a that the positive dark trion has negative $P_B$ and the negative dark trion has positive $P_B$, while the dark exciton has $P_B$ close to zero as the PL intensity from the two branches are similar. The valley information granted by the additional electron or hole can be understood by the different intervalley scattering mechanism. For doped samples under weak optical excitation, the electron-hole recombination, which generates the emitted photon, is determined by the optically excited minority carrier (hole for n-doped sample and electron for p-doped sample). For the case of the electron-doped sample, optically excited hole determines the electron-hole recombination. It has been shown[52] that the intervalley scattering is less likely and the hole prefers to stay in the same valley as it is excited, say K valley (Fig. 3c), and a photon will be emitted as a result of the e-h recombination in the K valley. As a result, the PL from the negative dark trion will exhibit the same valley polarization of the PL from the bright exciton. In the case of the p-doped sample, the e-h recombination will be determined by the optically excited electron. However, the intervalley scattering of electrons is much faster. In particular, the electron in the 2$^{nd}$ conduction band (shown in Fig. 3b) could be more likely to scatter to the other valley due to efficient coupling to the K point phonons, compared with the possibility of relaxing to the 1$^{st}$ CBM in the same valley as it requires flipping the spin. As a result, the emitted photon will arise from the e-h recombination in the other valley (K' in Fig. 3b) which has the opposite valley polarization as the bright exciton. In the presence of the out-of-plane magnetic field, the energy degeneracy between K and K' valley is lifted, and the valley information of the emitted photon can be resolved in the magneto-PL spectra as the two well-resolved branches (Fig. 2a and 2b). The higher valley polarization of the negative dark trion, stemming from the prohibitive intervalley scattering of the hole, making it an ideal candidate for valleytronic applications.

Considering the complexity of valley polarization associated with the dressing of free carriers and especially the weak PL from the negative dark trion, the "cross" pattern could be hard to resolve, especially for the negative dark trion with relatively weak PL (see Supplementary Note 2 and 3). Therefore, direct evidence of the dark trion is needed. Since the dark trions also radiate through the out-of-plane dipole as the dark exciton, the radiation pattern of the dark trion, if can be resolved, will be the unambiguous evidence of the dark trion. We achieved so by directly imaging the back focal plane[53], which is a Fourier transform of the real space image and has been used to resolve the energy dispersion in momentum space for photonic cavity[54,55]. Recently it has been applied to resolve the exciton radiation pattern of the 2D semiconductor[30,56]. Here we resolve the spectra information of the PL using the horizontal array of the pixels of the silicon charge-coupled device (CCD) and resolve the radiation angle, defined here as the angle from the normal, using the vertical array of the CCD camera (schematic setup in Supplementary Note 4).

The spectral-resolved and radiation-angle-resolved experimental data at 4.2 K are shown in Fig. 4. It is evident that the radiation pattern of the dark trion (Fig. 4a and 4c) is distinctively different from other excitonic complexes such as the bright exciton, but the same as that of the dark exciton (Fig. 4b). The maximum PL intensity is for the dark trions or the dark exciton is located at ~ 64 degrees, limited by the N.A. (~ 0.9) of our objective. In contrast, the PL intensity of the bright exciton is maximized near zero degree. The direct observation of the radiation pattern thus provides a decisive determination of the dark trions.

Finally, the well-resolved different excitonic peaks allow us to reveal the dynamics of the positive dark trion through time-resolved PL (TRPL) measurement directly[42]. Fig. 5a shows the PL spectrum obtained with a pulse laser excitation at 1.908 eV at 42 K, with the excitation power of 50 µW and the gate voltage of -2.5 V, and $X_D^+$ and $X^+$ are well-resolved. Since both $X_D^+$ and $X^+$ are charge-neutral excitons bound to a free hole, we assume the dielectric environment is similar at the same gate voltage, and we can directly compare the lifetimes of these two types of p-trions. The TRPL spectra (dots) for $X_D^+$ and $X^+$ are presented in Fig. 5b, and the lifetimes are extracted by the convolution of the instrument response function (IRF) (solid lines) with a single exponential function $I = Ae^{-t/\tau}$ for $X^+$ and a biexponential function $I = A_1 e^{-t_1/\tau_1} + A_2 e^{-t_2/\tau_2}$ for $X_D^+$, respectively[57–60]. The results indicate that the lifetime of $X^+$ only shows a fast component while that of $X_D^+$ exhibits both a dominant fast component and a slow component (see Supplementary Note 4 for a detailed analysis). We attribute the slow component to the contribution from the residue of the possible slowly-decaying PL from defects and thus use the fast component as the dark trion lifetime to compare with that of $X^+$. As shown in Fig. 5b, at the gate voltage of -2 V, the lifetime of $X_D^+$ (~ 90 ps) is more than 10 times longer than that of $X^+$ (~ 5 ps), which confirms that the positive dark trion inherits the long lifetime from the dark exciton[20,22,25,37]. The lifetime of the dark trion is also a sensitive function of the gate voltage. As shown in Fig. 5c, the lifetime of $X_D^+$ can be tuned from ~ 215 ps to ~ 50 ps as we change the gate voltage from -0.5 V to -4 V, which is expected as the increased hole doping increases the non-radiative channels and decreases the positive dark trion lifetime. The long lifetime of ~ 215 ps in the slight n-doping WSe$_2$ is comparable to that of the charge-neutral dark exciton (~ 250 ps) reported previously[42].

In summary, we have directly observed the dark trions through the back focal plane imaging of the radiation pattern of BN encapsulated monolayer WSe$_2$. The binding energy is determined to be ~ 15 meV. The large g-factor (~ -9) of the dark trions can be utilized to break valley degeneracy through an external magnetic field or proximity field effects[61,62]. The discovery of the dark trions illustrates a new way to tune the dark exciton complexes in WSe$_2$ through electrostatic gating. Dressing the dark exciton with the additional free carrier encodes the valley information into the otherwise valley-depolarized dark exciton. In addition, the lifetime of the dark trion can be sensitively changed from ~ 215 ps to ~ 50 ps by the control of doping. The dark trions thus open the door to new possibilities of valleytronic and excitonic applications.

**Acknowledgment**

We thank Dr. Lei Shi and Shengnan Miao for helpful discussions. Z. Li and S.-F. Shi acknowledge support from AFOSR through Grant FA9550-18-1-0312. T. Wang and S.-F. Shi acknowledge support from ACS PRF through grant 59957-DNI10. Z. Lian and S.-F. Shi acknowledge support from NYSTAR through Focus Center-NY–RPI Contract C150117. S. Tongay acknowledges support from NSF DMR-1552220 and DMR 1838443. The device fabrication was supported by Micro and Nanofabrication Clean Room (MNCR) at Rensselaer Polytechnic Institute (RPI). K.W. and T.T. acknowledge support from the Elemental Strategy Initiative conducted by the MEXT, Japan and the CREST (JPMJCR15F3), JST. Z. Lu. and D.S. acknowledge support from the US Department of Energy (DE-FG02-07ER46451) for magneto-photoluminescence measurements performed at the National High Magnetic Field Laboratory, which is supported by National Science Foundation through NSF/DMR-1644779, and the State of Florida. S.-F. Shi also acknowledges the support from a KIP grant from RPI.

**Supporting Information:**

The Supporting Information is available free of charge on the ACS Publications website:

Additional information on the gate-voltage and magnetic field dependent PL, radiation pattern detection technique and the detail of extracted lifetime from convolution (PDF)

**Competing financial interests:** The authors declare no competing interests.

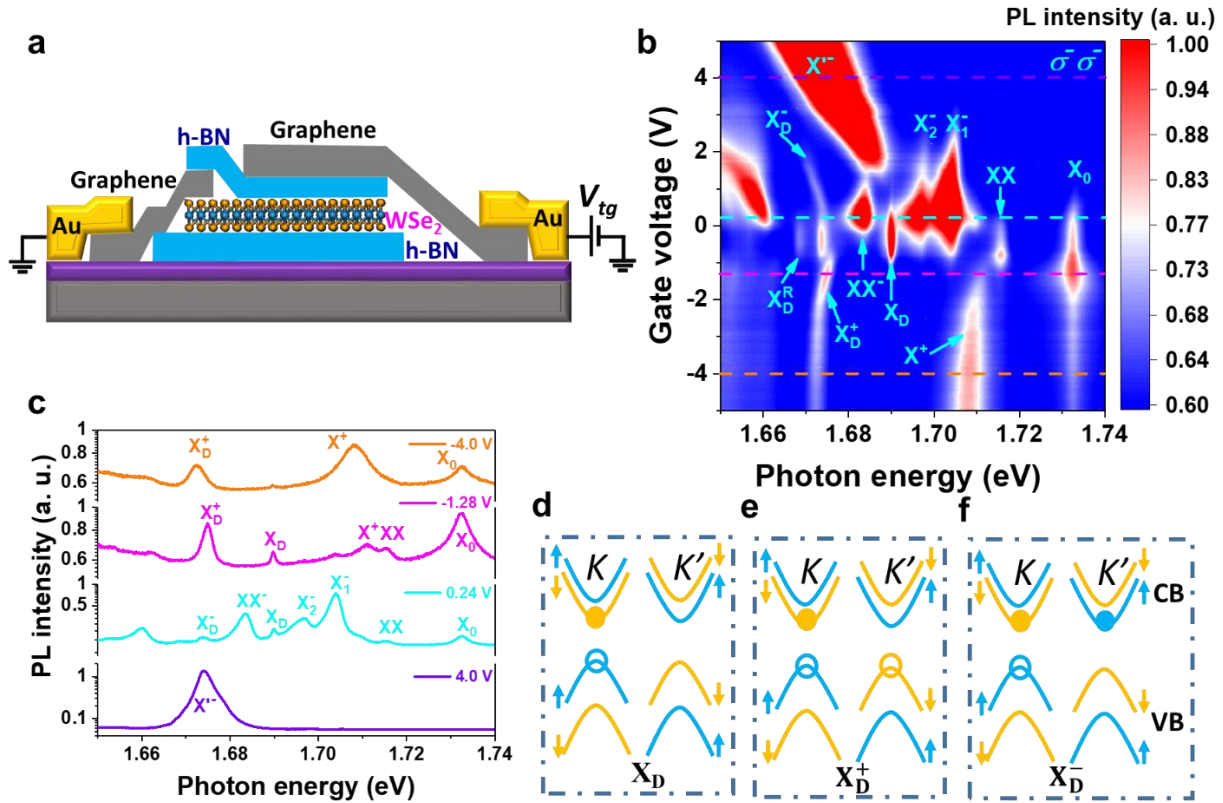

**Figure 1. Gate-dependent PL spectroscopy of the BN encapsulated monolayer WSe$_2$ device.** (a) Schematic representation of the BN encapsulated monolayer WSe$_2$ device. One piece of few-layer graphene is used as the electrode and another piece is used as the top gate electrode. (b) PL spectra at 4.2 K as a function of the top gate voltage for the BN encapsulated monolayer WSe$_2$ device. The color represents the PL intensity. The excitation is a CW laser centered at 1.879 eV with an excitation power of 40 μW. (c) PL spectra at the gate voltages of -4.0 V, -1.28 V, 0.24 V and 4.0 V, respectively, corresponding to the dash lines in (b). (d-f) are schemes of the dark exciton ($X_D$), positive dark trion ($X_D^+$) and negative dark trion ($X_D^-$), respectively.

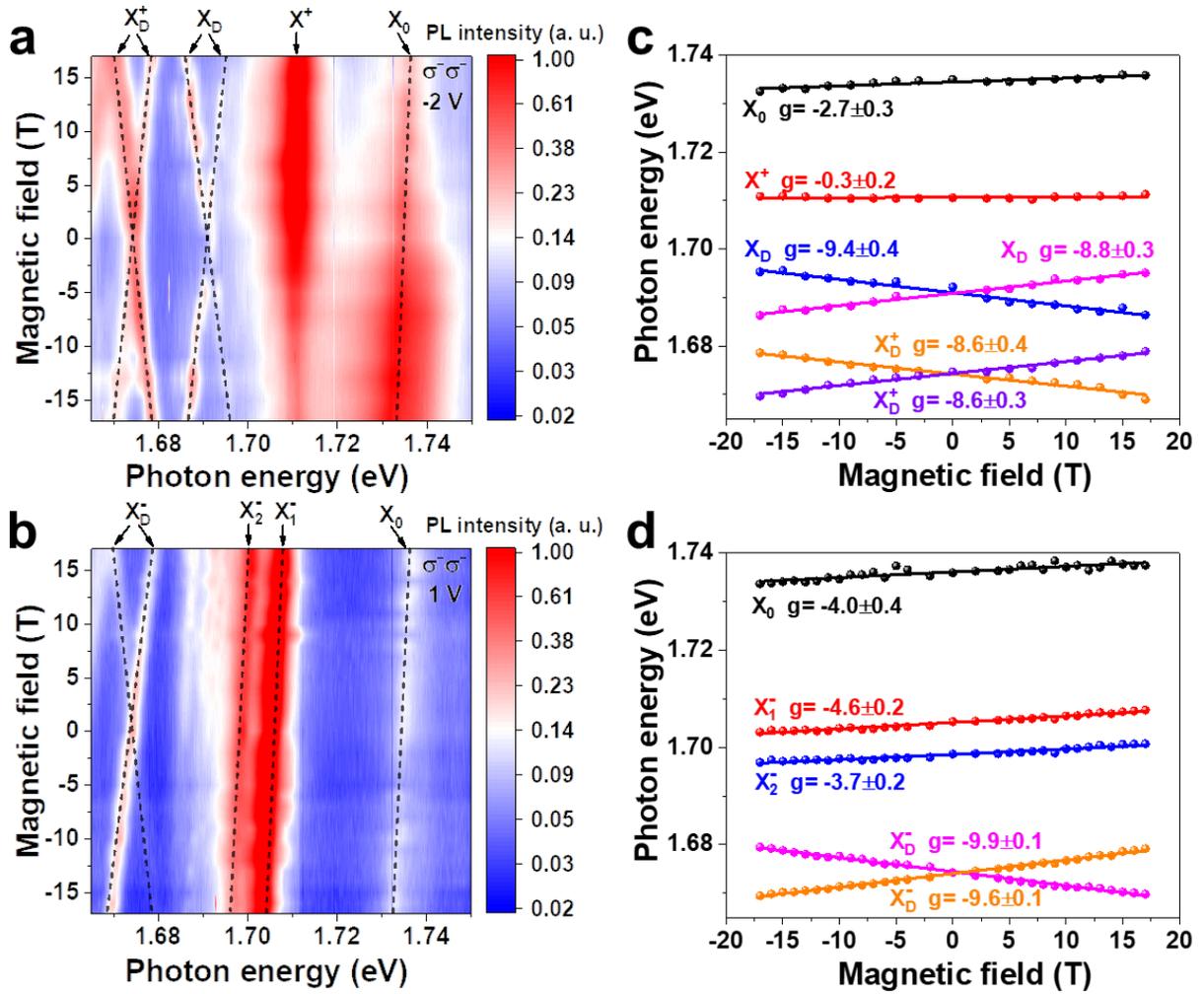

**Figure 2. Gate-dependent magneto-PL spectroscopy of monolayer WSe$_2$.** (a-b) are color plots of the PL spectra as a function of the magnetic field for the gate voltage of -2 V (hole-doped) and 1 V (electron-doped), respectively. (c-d) are the PL peak shift for different excitonic complexes extracted from (a) and (b). The Zeeman shift of each peak is utilized to calculate the associated g-factor through a linear fitting, for the gate voltages of -2.0 V and 1.0 V, respectively.

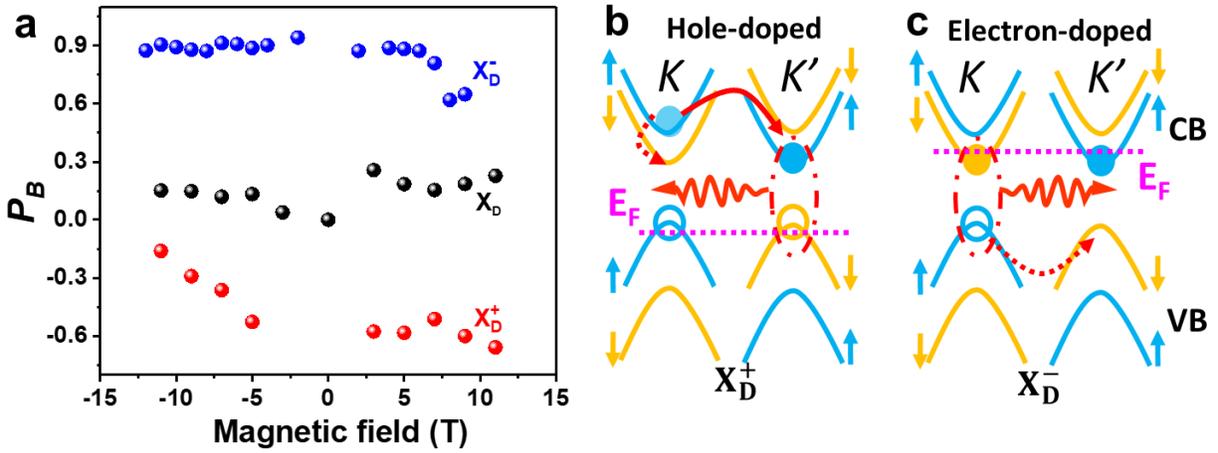

**Figure 3. Valley polarization under the magnetic field.** (a) the valley polarization $P_B$ of negative dark trion, dark exciton and positive dark trion under magnetic field, $P_B = \frac{I(K')-I(K)}{I(K')+I(K)}\sigma(B)$, where $I$(K) and $I$(K') are the PL intensity from K and K' valley, respectively. $\sigma(B)$ is 1 or -1 for positive or negative B field, respectively. (b-c) are the schematics of electron and hole recombination for positive dark trion and negative dark trion, respectively.

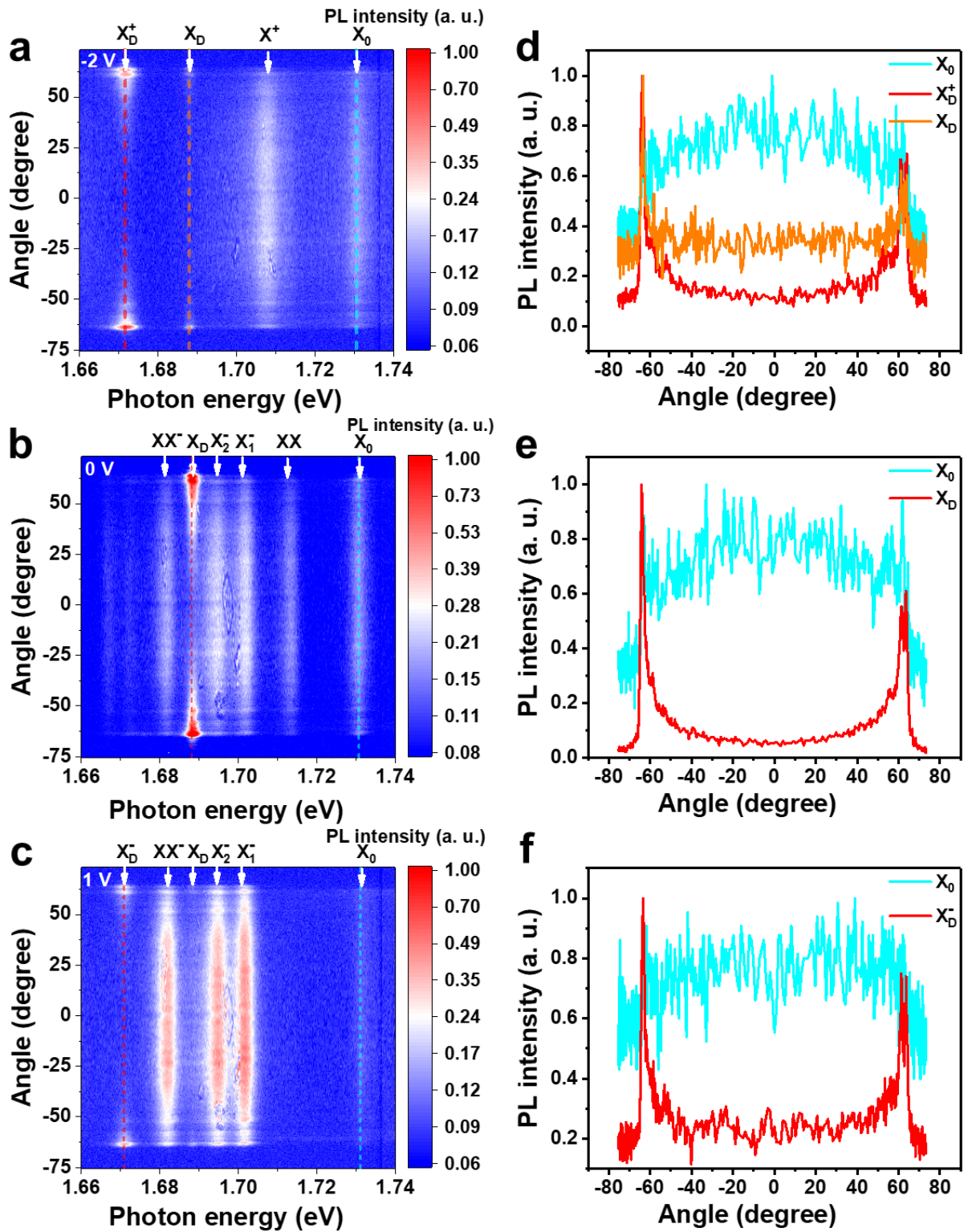

**Figure 4. Back focal plane imaging of the PL radiation pattern for different excitonic complexes.** (a-c) Color plots of PL intensity for different excitonic complexes as a function of emitted photon energy and the radiation angle. The radiation angle is resolved through imaging the back focal plane directly. (a), (b) and (c) are the experimental data

obtained for the top gate voltage of -2 V, 0 V, and 1 V, showing the radiation pattern for the positive dark trion, dark exciton and negative dark trion, respectively. (d-f) PL intensity as a function of the radiation angle (vertical line cut in a-c) of the positive dark trion, dark exciton and negative dark trion in the corresponding color plot, compared with the bright exciton ($X_0$) PL intensity at the same gate voltage.

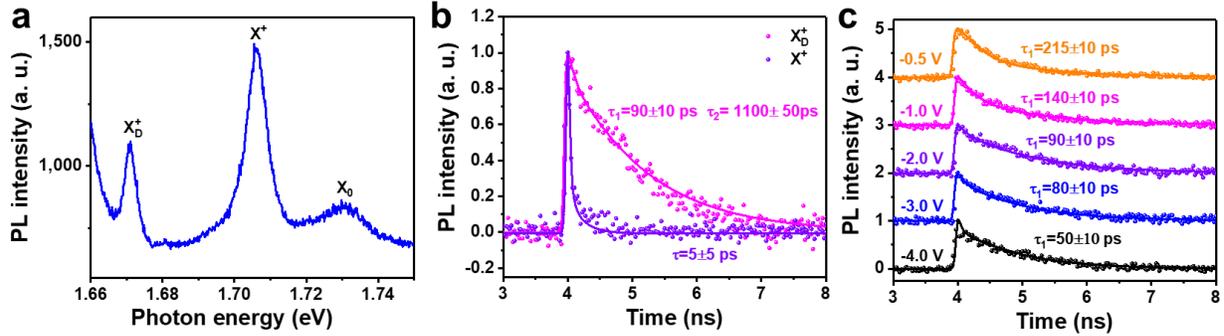

**Figure 5. Time-resolved PL spectra of $X_D^+$ of monolayer WSe$_2$.** (a) PL spectrum with a pulse laser excitation centered at 1.908 eV at 42 K, with the excitation power of 50 μW and the gate voltage of -2.5 V. (b) Comparison of the TRPL spectra of the positive bright trion (p-trion, purple) and positive dark trion (magenta) at the gate voltage of -2.0 V. The time-resolved PL data (dot) are convoluted (solid line) with IRF, using a single exponential function $I = Ae^{-t/\tau}$ for $X^+$ and a biexponential function $I = A_1 e^{-t_1/\tau_1} + A_2 e^{-t_2/\tau_2}$ for $X_D^+$. (c) TRPL spectra of the positive dark trion as a function of the gate voltage (dots), with the extracted lifetime for the fast component shown.

TOC

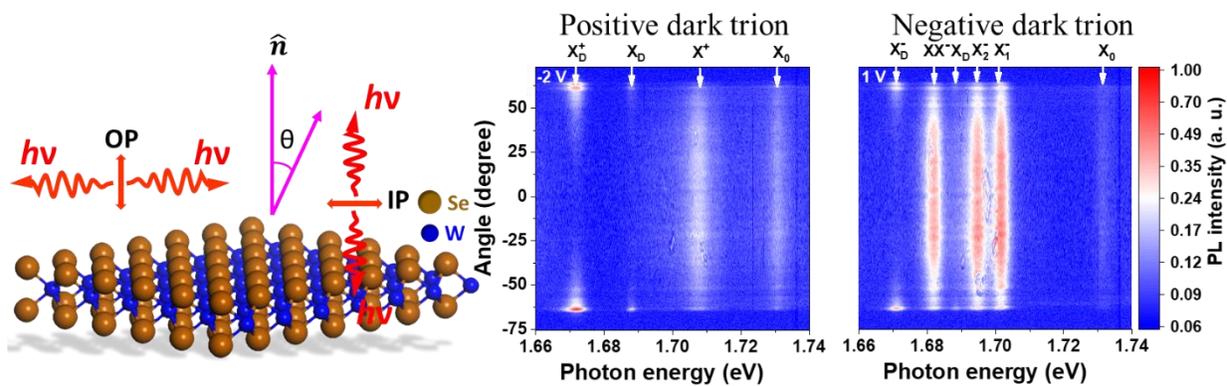